\documentclass[prd,nofootinbib,twocolumn,superscriptaddress,preprintnumbers,balancelastpage,nofootinbib]{revtex4-2}

\usepackage{graphicx}  % needed for figures
\usepackage{dcolumn}   % needed for some tables
\usepackage{bm}        % for math
\usepackage{amssymb,amsmath}
\usepackage{array}
\usepackage{slashed}   % for math
\usepackage{multirow}
\usepackage{tabularx}
\usepackage{booktabs} 
\usepackage{multirow}
\renewcommand{\eqref}[1]{Eq.~(\ref{#1})}

\usepackage{dsfont}
\usepackage{comment}
\usepackage{physics}
\usepackage{soul}

\usepackage[usenames,dvipsnames,svgnames,table]{xcolor}
\usepackage{silence}
\usepackage{capt-of}
\usepackage[colorlinks=true,allcolors=black]{hyperref}

\graphicspath{{./}{../}{body/}{../body/}{images/}{../images/}}

\def\be{\begin{equation}}
\def\ee{\end{equation}}
\def\bea{\begin{eqnarray}}
\def\eea{\end{eqnarray}}

\hypersetup{
    colorlinks=true,
    urlcolor=blue,
    linkcolor=blue,
    citecolor=blue
}

\definecolor{cborange}{HTML}{e69f00}
\definecolor{cbgreen}{HTML}{009e73}
\definecolor{cbyellow}{HTML}{f1dd42}
\definecolor{cblblue}{HTML}{56b4e9}
\definecolor{cbblue}{HTML}{0000FF}
\definecolor{defgrey}{HTML}{808080}
\definecolor{defgreen}{HTML}{008000}
\definecolor{defred}{HTML}{FA5F55}

\def\beq{\begin{equation}}
\def\eeq{\end{equation}}

\def\be{\begin{eqnarray}}
\def\ee{\end{eqnarray}}

\begin{document}
\widetext

\title{The Equivalence Principle in the Dark Sector in light of DESI}
%\title{Dark matter self-interactions before and after DESI}
\author{Andrea Clini}\email{andrea.clini@unimi.it}
\author{Pierre Zhang}\email{pierre.zhang@unimi.it}
\author{Emanuele Castorina}\email{emanuele.castorina@unimi.it}
\author{Maria Archidiacono}\email{maria.archidiacono@unimi.it}
\affiliation{Dipartimento di Fisica “Aldo Pontremoli”, Universit\`a degli Studi di Milano and \\INFN, Sezione di Milano,
Via Celoria 16, 20133 Milan, Italy}

\begin{abstract}
Recent measurements of the distribution of galaxies by the Dark Energy Spectroscopic Instrument (DESI) are in mild tension with a standard $\Lambda$-Cold Dark Matter cosmological model inferred from the anisotropies of the Cosmic Microwave Background (CMB) when extrapolated to low redshifts.
Several solutions to explain this discrepancy in terms of new physics have been put forward, notably from new dynamics in a complex Dark Sector.
Here we report on a resolution to this discrepancy that postulates that the inertial and gravitational masses of dark matter particles are not the same. 
Our best fit to CMB data combined either with Baryon Acoustic Oscillation (BAO) and Full Shape (FS) data from DESI DR1 (Data Release 1), or with BAO DR2 data alone, suggests the presence of a new long-range force in the Dark Sector with strength of $\beta \sim 0.5\% $ relative to gravity, at the $\sim 2.5\sigma$ level. 
A preliminary analysis combining FS DR1 and BAO DR2 data confirms the statistical power of DR2 and raises the evidence for dark forces to $\sim 3\sigma$, albeit ignoring the not yet available cross-correlation between the two datasets.
\end{abstract}

\maketitle

%===========================================================
\noindent
\section{Introduction}\label{sec:intro}
%==============================================================
The existence of Dark Matter (DM) is required to explain a large variety of gravitational phenomena, from the anisotropies of the Cosmic Microwave Background (CMB) produced roughly $380000$ years after the Big Bang, to the large-scale distribution of galaxies we observe today, the so-called Large Scale Structure (LSS) of the Universe.
In a minimal standard $\Lambda$-Cold Dark Matter ($\Lambda$CDM) cosmological model, DM is a highly non-relativistic species, hence the attribute cold, with no interaction with other particles, either in the visible or in a possible extended Dark Sector.
This also includes any long-range force between dark matter particles other than gravity, thus enforcing the validity of the Equivalence Principle (EP) in the Dark Sector.

Recently, the Dark Energy Spectroscopic Instrument (DESI) second data release (DR2) provided new measurements of the expansion rate of the Universe at low redshifts, in the form of angular diameter and comoving distances relative to the sound horizon imprinted in the distribution of baryons at the time of recombination~\cite{2503.14738}.
The best-fit $\Lambda$CDM model to DESI DR2 is found to be in tension with the corresponding standard model inferred from the analysis of CMB experiments such as \textit{Planck} \cite{1807.06209}, the Atacama Cosmology Telescope (ACT) \cite{2503.14452} and the South Pole Telescope (SPT) \cite{2506.20707}.
These findings are consistent with previous DESI DR1 measurements of both the expansion rate and the clustering of galaxies~\cite{2411.12021,2411.12022,DESI:2024lzq,DESI:2024uvr}. 
Such a discrepancy, ranging at the $\sim 2-3\sigma$ levels depending on the specific combination of datasets, has recently received considerable attention.
Several proposals to ameliorate the tension have focused on deviations of Dark Energy from a Cosmological Constant \cite{2411.12022,2503.14743,2405.13588,2404.08056,2404.19437,2405.03933,2407.15832,2409.17019,2503.04602,2504.06118,2504.07679,2507.07193,2507.16970}, possibly violating the null energy condition, non-zero spatial curvature \cite{2505.00659} and many others \cite{2504.16932,2504.21813,2407.18252,2503.16415,2505.05450,2507.03090,2507.13925,2510.14957,2605.18716,Teixeira:2024qmw,Li:2026xaz}.
One way to look at the tension between DESI and CMB data is that the former prefers a lower value of today's density of matter $\Omega_m=\omega_m/h^2$, which in a $\Lambda$CDM model this typically implies that the Hubble constant $H_0$ is higher in DESI data than it is in a CMB analysis. 

A lower abundance of matter today compared to what inferred from early-Universe data could be the result of new dynamics in the Dark Sector.
DM could decay over cosmological timescales, though this possibility seems to be currently excluded \cite{2603.03284, Lynch:2025ine}.
Alternatively, if DM particles feel a new long-range force mediated by an ultra-light scalar field, effectively violating the EP in the Dark Sector, the DM mass and thus its energy density decrease faster than the traditional volume scaling of $(1+z)^3$, where $z$ is the cosmological redshift \cite{Archidiacono:2022iuu,Bottaro:2023wkd}.
This is the unavoidable result of the transfer of energy from the DM field to the mediator as more interactions take place over time.

The goal of this work is to assess whether a pure dark fifth force \cite{Archidiacono:2022iuu}, \emph{i.e.} of infinite range, can alleviate the tension between DESI and CMB datasets.
Our main result (cf. Fig.~\ref{fig:baseline_bound}) is that a new long-range interaction, that interacts with the totality of DM, of strength $\beta \sim 5 \text{‰}$ of gravity provides a statistically significant, at the $\sim 2.5\sigma$ level, solution to this discrepancy.
This reconciliation is suggestively illustrated in Fig.~\ref{fig:BAO}, where the fifth-force best fit improves the agreement between CMB-inferred cosmology and LSS measurements.
Our findings confirm previous analyses \cite{2407.18252,Costa:2025kwt} based on background distances, but with the important addition of the full shape of the power spectrum of DESI galaxies, thus providing a non-trivial test of the model at both the background and perturbation level.

The rest of this paper is organized as follows. 
In Sec.~\ref{sec:theory}, we review the main physical features of a Universe in the presence of a dark fifth force, focusing on the effects in cosmological observables. 
Next, we confront in Sec.~\ref{sec:results} this setup to various combinations of CMB and LSS data, testing whether the EP holds in the Dark Sector. 
Finally, we conclude in Sec.~\ref{sec:conclusions}.
Additional materials and technical details are relegated in appendices. 

\begin{figure*}
    \centering
    \begin{minipage}[t]{0.49\textwidth}
        \vspace{0pt}
        \centering
        \includegraphics[width=\linewidth,height=0.42\linewidth,keepaspectratio]{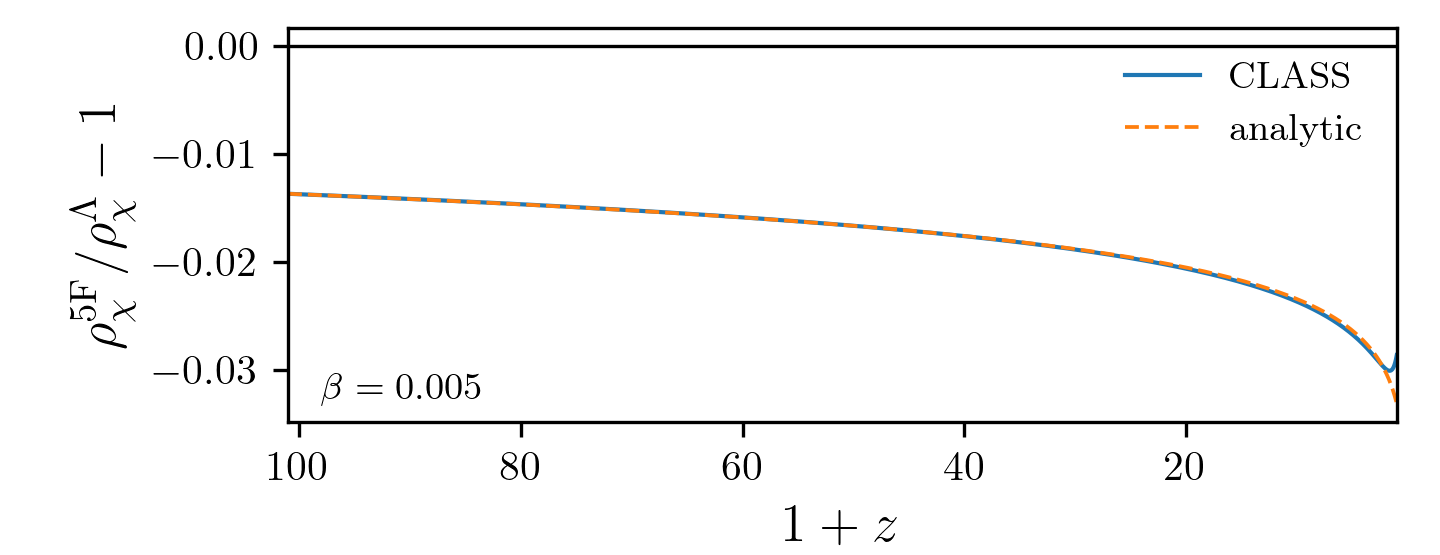}
        \vspace{0.4em}
        \includegraphics[width=\linewidth,height=0.42\linewidth,keepaspectratio]{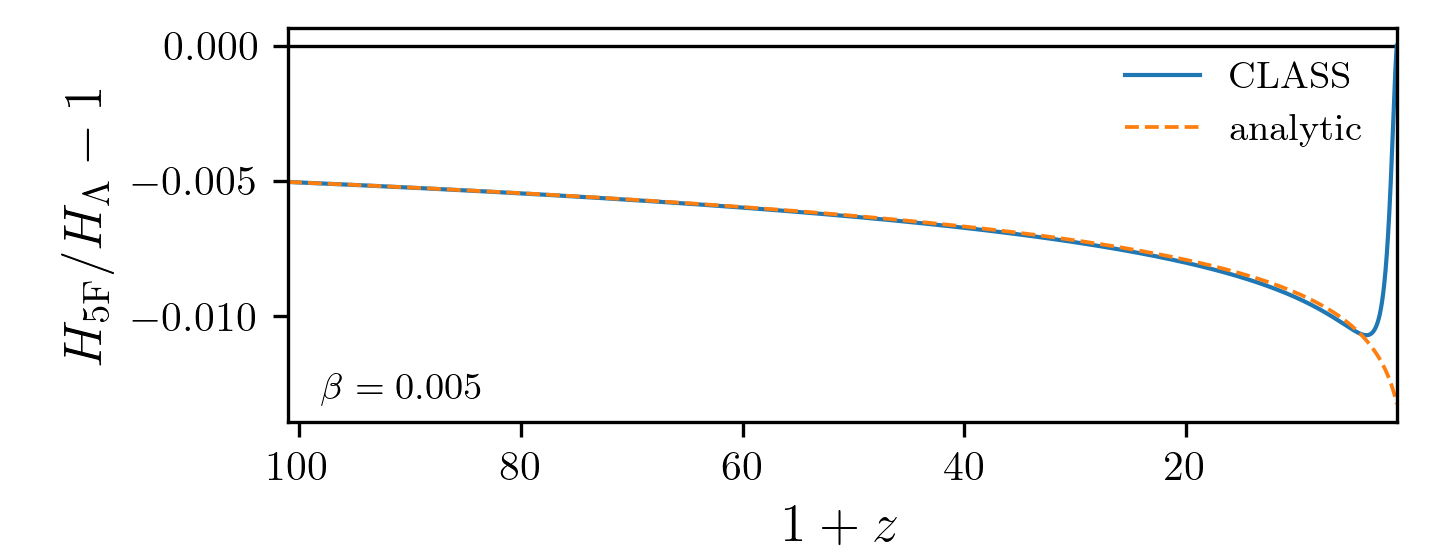}
    \end{minipage}
    \hfill
    \begin{minipage}[t]{0.49\textwidth}
        \vspace{0pt}
        \centering
        \includegraphics[width=\linewidth,height=0.84\linewidth,keepaspectratio]{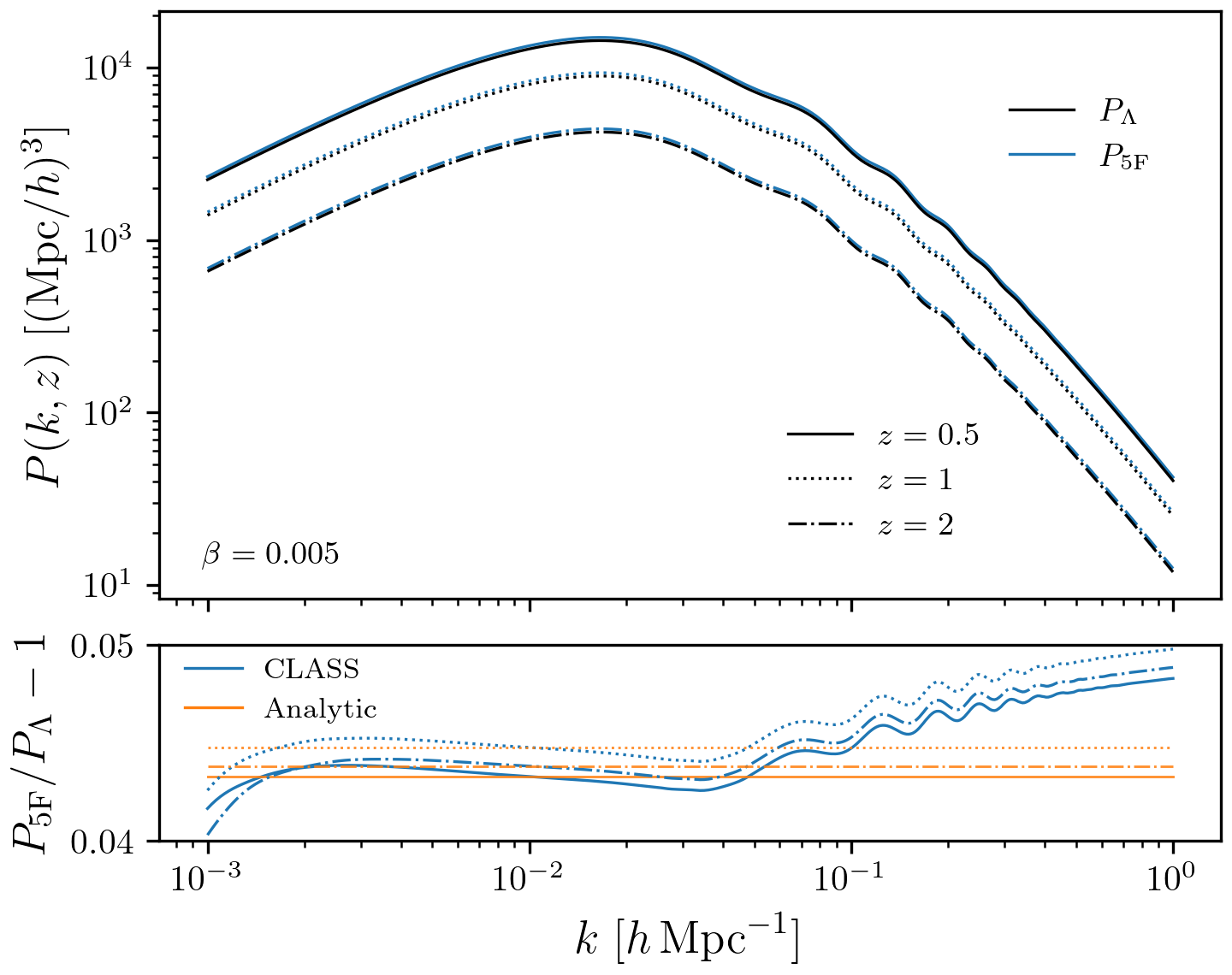}
    \end{minipage}

    \vspace{-8pt}
    \caption{Deviations from $\Lambda$CDM by the presence of a dark fifth force with  strength $\beta = 0.5\%$ relative to gravity on \textit{(i)} background quantities -- DM energy density $\rho_\chi(z)$ and Hubble parameter $H(z)$  as a function of redshift $z$ (\textit{left panels}); and \textit{(ii)} the linear matter power spectrum $P(k, z)$ as a function of wavenumber $k$ for $z = 0.5, 1$, and $2$ (\textit{right panel}). 
    The numerical solutions (\textit{blue curves}) from the Boltzmann code \texttt{CLASS}  are compared to the approximate, leading log-enhanced analytical solutions (\textit{orange curves}) presented in the main text. 
    Here as in all subsequent cosmological analyses presented in this work, we assume that all the dark matter interacts with the fifth force.}
    \label{fig:theory}
\end{figure*}

%===========================================================
\section{Background and perturbation dynamics}\label{sec:theory}
%==============================================================

The physical effects of a violation of the EP in the Dark Sector can be realized by DM particles interacting with a light mediator whose Compton wavelength $\sim 1/m$ is larger than the present Hubble radius, thus effectively acting as a new long-range force. 
This scenario can be consistently included in Boltzmann cosmological codes like \texttt{CLASS}, and we refer the reader to Refs.~\cite{Archidiacono:2022iuu,Bottaro:2023wkd} for more technical details. 
Here, we limit ourselves to the most general features of the model, with an eye on its potential implication at resolving the cosmological tension between CMB and LSS data touched upon in Sec.~\ref{sec:intro}. 

We assume the new force is mediated by a light scalar field $\varphi$ with mass $m_\varphi \lesssim H_0$, interacting with a scalar DM $\chi$ via $\mathcal{L}_{\rm int} = - \kappa \varphi \chi^2$.
Upon redefining the light mediator via $s \equiv G_s^{1/2} \varphi$, with $G_s \equiv \kappa ^2 / m_\chi^4$, the part of the Lagrangian quadratic in $\chi$ becomes $\tfrac{1}{2} m_\chi^2(s) \chi^2$, where $m_\chi^2(s) = m_\chi^2 (1+2 s)$, showing how a violation of the EP leads to a space-time dependent mass for the DM.
The ratio between the new force and gravity is then parameterized as $\beta \equiv G_s /(4 \pi G_N)$. 

The background equations of motion for the DM energy density $\rho_\chi$  and the mediator $s$ read
\begin{align}
    &\rho_\chi' + 3 \mathcal{H} \rho_\chi  = \rho_\chi \widetilde{m}_s s'\,, \\
    & s'' + 2 \mathcal{H}s' + a^2 (m_\varphi^2 s + 4 \pi G_N \beta \widetilde{m}_s \rho_\chi) =0\,,
\end{align}
where derivatives are with respect to conformal time, $\mathcal{H} = a'/a$ with $a$ the scale factor, and $\widetilde{m}_s := \dd \log m_\chi(s)/ \dd s$.
The DM density admits the formal solution $\rho_\chi =  n_\chi \, m_\chi(s)$, with the DM number density scaling as $n_\chi \propto a^{-3}$, which in turn allows us to solve for $s$ to leading order in $\beta$.
The final result, during matter domination, and in comparison with the corresponding $\Lambda$CDM quantities, is 
\begin{align}\label{eq:bkg_rho_chi}
  &  \frac{\rho_\chi}{\rho_{\chi}\big|_{\Lambda\rm{CDM}}} \simeq 1- \beta f_\chi \widetilde{m}_s^2 \log \frac{a}{a_{\rm eq}}\,,
   \\
& \label{eq:bkg_hubble} \frac{\mathcal{H}}{\mathcal{H}\big|_{\Lambda\rm{CDM}}} \simeq 1- \frac{\beta}{2} f_\chi^2 \widetilde{m}_s^2 \log \frac{a}{a_{\rm eq}}\,,
\end{align}
where $a_{\rm eq}$ is the scale factor at matter-radiation equality, and $f_\chi$ is the fraction of interacting DM with respect to the \emph{total} non-relativistic component, \emph{i.e.} including baryons.
As anticipated in the Introduction, both the matter density and the Hubble parameter are reduced with respect to the corresponding quantities in the standard cosmological model. 

At the level of perturbations, the physics during radiation domination is mostly unchanged compared to a $\Lambda$CDM scenario. 
For small $\beta$ we can thus neglect modifications to the physics prior to hydrogen recombination.
In particular, the value of the sound horizon of the baryon-photon fluid is unaffected by the presence of a new force.
This implies that $H_0$ and $\beta$ will be \emph{positively} correlated in any analysis involving CMB data, since, in order to keep the angular size of the sound horizon fixed to its measured value, the Hubble constant $H_0$ must increase to compensate for the dynamical effect of the new force in \eqref{eq:bkg_hubble}. 
Likewise, \eqref{eq:bkg_rho_chi} implies that $\Omega_m h^2$ and $\beta$ will be \emph{negatively} correlated in a CMB analysis, since early-time matter density determined prior to recombination should remain approximately fixed as $\beta$ changes. 
Hence, from the two arguments above, $\Omega_m$ and $\beta$ will be negatively correlated as well, which suggests that a non-zero $\beta >0$ can accommodate a lower abundance of matter at low redshifts.   
Together with the discrepancy between CMB and DESI data mentioned in the Introduction, these observations motivate to test the presence of an extra long-range force in the dark sector on the cosmological data, which we carry in the next section. 

Given the attractive nature of the new force, the growth of perturbations is enhanced with respect to a Universe where only gravity is present. 
For the perturbations in the total matter density $\delta_m$ and velocity divergence $\theta_m$, one finds an approximately scale-independent enhancement during matter domination \cite{Archidiacono:2022iuu} (see also Ref.~\cite{Costa:2025kwt}), 
\begin{align}
    \frac{\delta_m^{\rm lin}}{\delta^{\rm lin}_{m} \big|_{\Lambda\rm{CDM}}} & \simeq 1 + \beta f_\chi^2 \widetilde{m}_s^2 \log \frac{a}{a_{\rm eq}}\,  , \label{eq:perturbations}  \\
    \frac{\theta_m^{\rm lin}}{\theta^{\rm lin}_{m} \big|_{\Lambda\rm{CDM}}} & \simeq 1 + \frac{\beta}{2} f_\chi^2 \widetilde{m}_s^2 \log \frac{a}{a_{\rm eq}}  \, . \label{eq:v_perturbations}
\end{align}

The extra factor of $f_\chi$ in \eqref{eq:bkg_hubble}-\eqref{eq:v_perturbations} compared to the background evolution \eqref{eq:bkg_rho_chi}, is simply due to the fact that the interacting DM contributes by $f_\chi$ to the total matter density.

Finally, an unbiased analysis of LSS data requires the extension of nonlinear perturbation theory in the presence of the new force.
This task was carried out in Ref. \cite{Bottaro:2023wkd}, and we refer the reader to that work for an extended discussion.
The bottom line is that for small values of $\beta$, the nonlinear time evolution is well described by the standard treatment as in $\Lambda$CDM, since this accounts for all $\beta \log (a/a_{\rm eq})$ corrections directly inherited from the modification of the linear growth in~\eqref{eq:perturbations} and~\eqref{eq:v_perturbations}, with negligible non-log-enhanced $\mathcal{O}(\beta)$ errors in the higher-order perturbative solutions to the equations of motion. 

Figure~\ref{fig:theory} compares the analytic expressions \eqref{eq:bkg_rho_chi}-\eqref{eq:perturbations} with the numerical solutions from \texttt{CLASS}~\cite{Blas:2011rf,Lesgourgues:2011re,Archidiacono:2022iuu} for a fifth-force Universe with  $\beta = 0.005$ and all other cosmological parameters fixed to \textit{Planck} best-fit values \cite{1807.06209}, relative to a corresponding $\Lambda$CDM model. 
Here and in the rest of this work, we assume that all DM is interacting and a flat Universe, so that the fraction of interacting DM to total matter is simply $f_\chi \simeq \Omega_\chi / \Omega_m \sim 0.85$. 
Throughout, we also assume  that the scalar mediator has a negligible initial displacement $s_{\rm ini} \simeq 0$, yielding $\tilde m(s) \simeq 1$, such that the size of the modifications by the presence of the new force are solely controlled by (and scale with) $\beta$. 
We see that our analytical estimates keeping only the log-enhanced corrections scaling as $\sim \beta \log(a/a_{\rm eq})$ agree well with the numerical solutions, validating the main expected behavior discussed above.

Although the difference in the shape of the matter power spectrum in Fig.~\ref{fig:theory} is mostly a constant offset, for the power spectrum of galaxies in redshift space a rescaling of the density amplitude can not be fully absorbed by a shift in the galaxy bias parameters. 
Mainly, redshift-space distortions, that involve correlations between both $\delta_m$ and $\theta_m$, break, already at the linear level, the degeneracies between amplitude and bias parameters. 
For $\beta \sim 0.5\%$, the enhancement of the power spectrum relative to $\Lambda$CDM is about $\sim 5\%$. 
Assuming that measurements of the galaxy power spectrum quadrupole, which primarily constrain the amplitude of the redshift-space distortions, have a comparable precision, this enhancement would be compensated by roughly a $1\sigma$ downward shift in $f \sigma_8$, the product of the growth rate $f$ with the clustering amplitude $\sigma_8$. 
We thus expect that measurements of the amplitude of the galaxy power spectrum multipoles will be complementary with measurements of the background expansion: 
For comparison, typical percent-level measurements of BAO distances are expected to constrain $\beta$ at the $\sim 0.2\%$ level as estimated via the modifications in the ruler $H \equiv  \mathcal{H}/a$ highlighted in~\eqref{eq:bkg_hubble}.

%====================================================
\section{Cosmological analysis}~\label{sec:results}
%====================================================

\paragraph{Datasets and inference setup}
In this work, we consider various combinations of the following CMB and LSS datasets: \textit{Planck} PR3 temperature and polarization power spectra \cite{1807.06209}, the combination of \textit{Planck} and ACT CMB lensing \cite{ACT:2023dou, ACT:2023kun, Carron:2022eyg} and galaxy clustering data from DESI. 
DESI has publicly released the full DR1 datasets, consisting of the post-reconstructed BAO data, the full shape (FS) of the pre-reconstructed galaxy power spectrum multipoles, and their cross-correlations \cite{2411.12020,2404.03000,2411.12021}, while at this stage only BAO measurements are available in their DR2 \cite{2503.14738}.
%We will use the shorthand FS1, BAO1 and BAO2 to refer to the full shape DR1, the BAO DR1 and the BAO DR2 datasets, respectively.

The BAO distance data provide, at the given effective redshift $z$ of a galaxy sample, the quantities $H(z)r_d$ and $D_M(z)/r_d$, where $D_M(z)$ is the comoving transverse distance and $r_d$ is the sound horizon at the baryon drag epoch, offering precise measurements of the background evolution of the Universe at low redshifts~\cite{2404.03000,2503.14738}.

The full-shape data consists in the measurements of the multipoles of the redshift-space galaxy power spectrum. 
Following the DESI collaboration \cite{2411.12021}, we include, for each of the 6 redshift bins in DESI DR1, the first two even multipoles and restrict the fit to wavenumbers $0.02 < k / (\,h\,{\rm Mpc}^{-1}) < 0.2$.
On top of the cosmological parameters, we fit for additional nuisance parameters accounting for our ignorance about galaxy formation and small-scale physics. 
Further details on the full-shape analysis are given in App.~\ref{sec:desi_dr1_full_shape}.

\begin{figure}[ht!]
    \centering
    \includegraphics[width=1.0\linewidth]{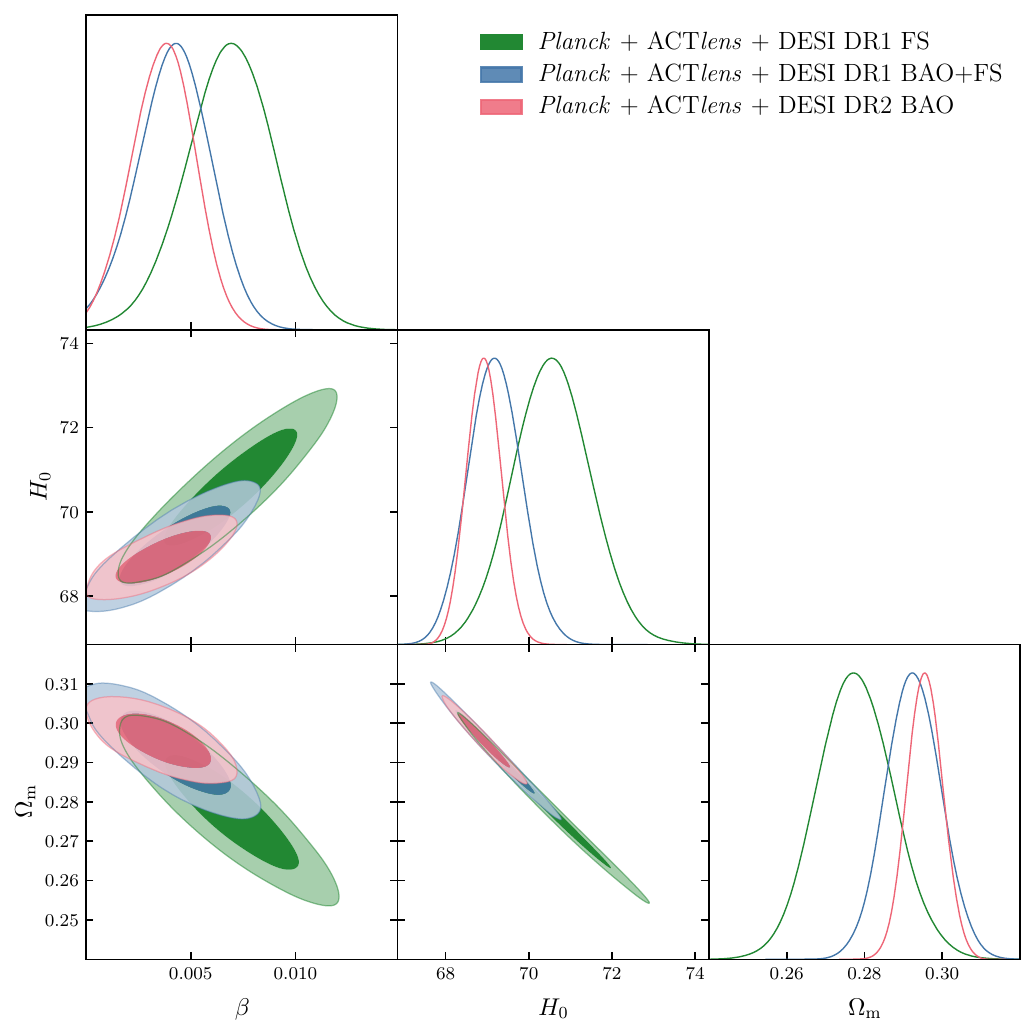}
    \caption{1D and 2D posterior distributions of the dark fifth force strength $\beta$ relative to gravity, the Hubble constant $H_0$, and the matter abundance $\Omega_m$ for the three baseline dataset combinations between CMB and LSS probes considered in this work. }
    \label{fig:baseline_bound}
\end{figure}

\begin{table*}[t]
\centering
\setlength{\tabcolsep}{18pt}
\begin{tabular}{lccc}
\toprule
    Dataset
    & $\beta$
    & $\Omega_{\rm m}$
    & $H_0 \, [\textrm{km}/s/\textrm{Mpc}]$ \\
\midrule
    \textit{Planck} + ACT\textit{lens}
    & $< 0.0152$ ($95\%$)
    &  $0.2858 \pm 0.0247$
    & $69.93 \pm 2.41$ \\

        + DESI DR1 BAO
    & $0.0042 \pm 0.0018$
    & $0.2934 \pm 0.0074$
    & $69.12 \pm 0.66$ \\

    + DESI DR1 FS
    & $0.0069 \pm 0.0021$
    & $0.2777 \pm 0.0097$
    & $70.56 \pm 0.93$ \\
    
    + DESI DR1 FS+BAO
    & $0.0042 \pm 0.0017$
    & $0.2926 \pm 0.0070$
    & $69.19 \pm 0.63$ \\

    + DESI DR2 BAO
    & $0.0037 \pm 0.0015$
    & $0.2956 \pm 0.0045$
    & $68.92 \pm 0.41$ \\

    + DESI DR1 FS + DR2 BAO
    & $0.0043 \pm 0.0014$
    & $0.2921 \pm 0.0042$
    & $69.22 \pm 0.39$ \\
    
    \cmidrule[0.05pt](lr){1-4}
    + DESI DR1 FS+BAO, free $\sum m_\nu$
    & $0.0046 \pm 0.0022$
    & $0.2925 \pm 0.0071$
    & $69.20 \pm 0.64$ \\

  %  + DESI DR1 FS + H0DN prior
  %  & $0.0097 \pm 0.0015$
  %  & $0.2635 \pm 0.0063$
  %  & $71.96 \pm 0.64$ \\
\bottomrule
\end{tabular}
\caption{$68\%$-credible intervals (mean and standard deviation) of representative cosmological parameters in a Universe with the presence of a fifth force, obtained from the various dataset combinations considered in this work. 
For \textit{Planck} + ACT\textit{lens} only, we show instead on $\beta$ the $95\%$-upper bound. 
Parameters not shown such as the baryon abundance $\omega_{\rm b}$ as mainly determined prior to recombination from CMB data remain practically consistent across all analysis setups. 
}
\label{tab:lrf_constraints}
\end{table*}

\begin{figure*}[ht!]
    \centering
    \includegraphics[width=0.95\textwidth]{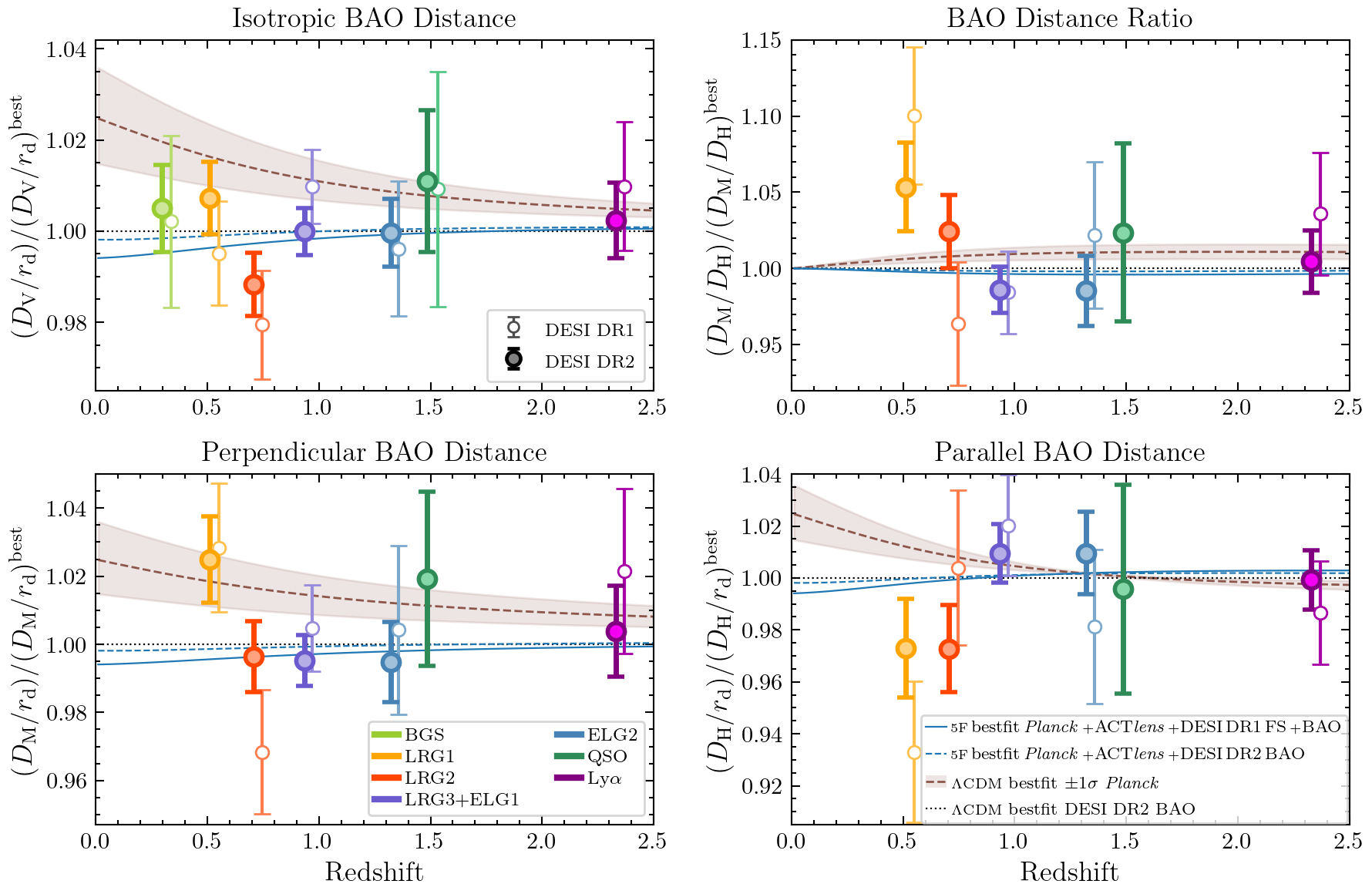}

    %\vspace{-9pt}
    \caption{BAO distance measurements from DESI DR1/DR2, compared to various best fits to the two baseline dataset combinations obtained in this work in the presence of a dark fifth force (5F). 
    $\Lambda$CDM best fits to data from CMB, LSS, or their combination, are shown for comparison. 
    The plots are normalized by the $\Lambda$CDM best fit to DESI DR2 BAO.}
    \label{fig:BAO}
\end{figure*}

\paragraph{Main results}
Our main results for our three baseline datasets are summarized in Fig.~\ref{fig:baseline_bound}.
Table~\ref{tab:lrf_constraints} reports the $68\%$-credible intervals on the cosmological parameters of interest from the various dataset combinations considered in this work. Triangle plots for the full set of cosmological parameters are shown in Fig.~\ref{fig:posterior_all} in App.~\ref{app:constraints}. 

Our baseline datasets, consisting of the three CMB+LSS combinations given by \textit{Planck} + ACT\textit{lens} + DESI DR2 BAO, \textit{Planck} + ACT\textit{lens} + DESI DR1 FS, and \textit{Planck} + ACT\textit{lens} + DESI DR1 FS+BAO,  indicate a preference for a non-zero fifth force between the $2.5$ and $3.5\sigma$, yielding $\beta = 0.0037 \, (0.0040) \pm 0.0015$, $\beta = 0.0069 \, (0.0070) \pm 0.0021$ at $68\%$, $\beta = 0.0042 \, (0.0049) \pm 0.0017$ at $68\%$CL, respectively, where the numbers in parenthesis are the maximum-a-posteriori values. 
The preference for a non-zero fifth force is further supported, albeit to a slightly lesser extent, by the difference in $\chi^2$ of the best-fit point relative to $\beta = 0$. 
For the baseline fits, we find $\Delta \chi^2 = -4.7$, $\Delta \chi^2 = -7.7$ and $\Delta \chi^2 = -5.9$, respectively, corresponding to a Gaussian-equivalent significance of $2.2\sigma$, $2.8\sigma$ and $2.4\sigma$ over $\Lambda$CDM, counting one extra degree of freedom.\footnote{Since $\beta$ is bounded from below at $\beta \equiv 0$,  Wilks theorem does not apply strictly, and we instead resort to Chernoff theorem that provides the asymptotic likelihood-ratio distribution for a single additional parameter whose null value lies on the boundary, for which the evidence reads $Z = \sqrt{|\Delta \chi^2|}$. }
As anticipated in Sec.~\ref{sec:theory}, since $\beta$ is positively correlated with $H_0$ and anti-correlated with $\Omega_m$, the non-zero positive value of $\beta$ drives the values of $\Omega_m$ down and of $H_0$ up, reflecting that the presence of a dark fifth force can lead to a better consistency between CMB and LSS data without the need to invoke any exotic form of dark energy. 
We however anticipate that, due to the comparatively large value in $\Omega_m$ preferred by the distance-redshift relation measured from supernova data~\cite{2511.07517}, further modeling assumptions beyond the scope of this work, such as the possibility that the mediator takes the role of a  quintessence field al late times~\cite{Archidiacono:2022iuu,2507.03090,Notari:2024rti,2503.16415,Gomez-Valent:2026ept}, will be required to bring full consistency with these additional low-redshift data.

For future reference, we further consider a mixed DESI combination in which DR1 FS data are supplemented with the latest DR2 BAO measurements. 
Although this analysis is not fully consistent, as the cross-correlation between the DR1 FS and DR2 BAO data is not yet publicly available (see however Ref.~\cite{2602.18761} suggesting that neglecting this small overlap does not significant impact the cosmological results),
it can serves, at the very least, a useful indication of the statistical power of the latest DR2. 
This indeed yields $\beta=0.0043\pm0.0014$, that, at face value, increases its determination by $\sim 20\%$ and reject the null hypothesis that the EP holds in the Dark Sector at the $ 3.1\sigma$ level (or $\sim 2.8\sigma$ from $\Delta \chi^2 = -8.0$). 

\paragraph{BAO distances}
To better understand how LSS data constrain the strength of the dark fifth force $\beta$, we compare in Fig.~\ref{fig:BAO} the DESI BAO measurements to the best fits obtained from our two baseline datasets. 
The presence of a non-zero fifth force restores concordance between the CMB-preferred cosmology with DESI low-redshift distance measurements. 
Since the CMB is practically blind to an interacting Dark Sector relative to the precision from LSS (see Tab.~\ref{tab:lrf_constraints}), a small non-zero $\beta \sim 0.005$ increases only slightly the (integrated) line-of-sight distance from us to the CMB, keeping the fit to CMB spectra mostly unaffected. 
In contrast, the relative distances measured in the low-redshift Universe are strongly impacted: the parallel and perpendicular distances to our line-of-sight appear to be smaller than in $\Lambda$CDM Universe, in line with reduction in ruler $\mathcal{H}$ showcased in~\eqref{eq:bkg_hubble}, which ends up to fit better the BAO measurements inferred from DESI data.

To sum up, new dark forces seem to alleviate the tension between CMB and LSS measurements. 
In light of our findings, we now report additional results based on extended cosmologies to assess the robustness and interpretation of the preference for $\beta>0$.

\begin{figure}[ht!]
    \centering
    \includegraphics[width=1.0\linewidth]{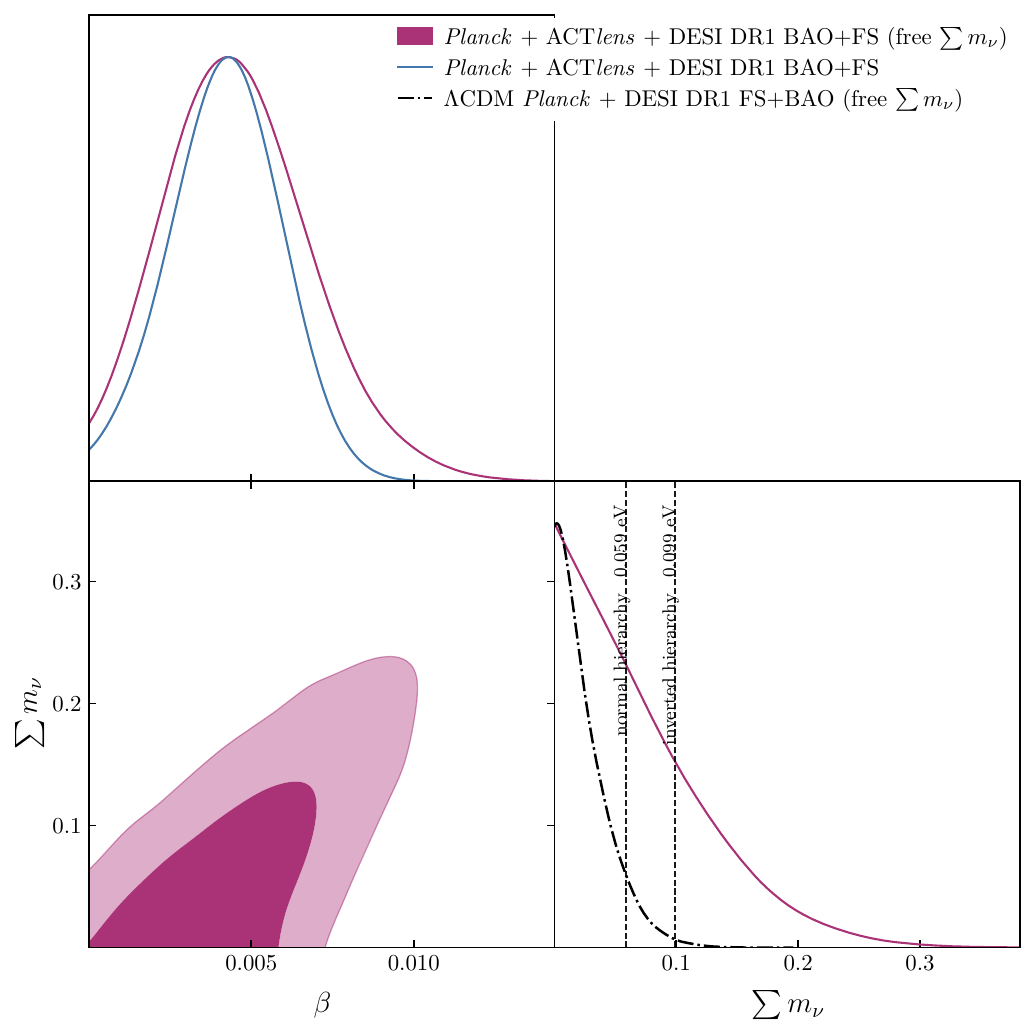}
    \caption{1D and 2D distributions on the dark fifth force strength $\beta$ and the sum of neutrino masses $\sum m_\nu$ from our fit to the baseline CMB + LSS data combination (\textit{red contours}). 
    In comparison, we show the 1D posterior of $\beta$ when fixing the total neutrino mass to minimal (\textit{blue contour}),
    the 1D posterior on $\sum m_\nu$ obtained in $\Lambda$CDM (\textit{black dashed contour}), and the minimal neutrino mass from oscillation experiments in normal and inverted hierarchy \cite{2410.05380} (\textit{dashed vertical lines}). }
    \label{fig:nus}
\end{figure}

\paragraph{Neutrino masses} 
Massive neutrinos that contribute to the total matter density budget free stream after decoupling from the Standard Model thermal bath, thereby suppressing the growth of structure at late times in comparison to a Universe with massless neutrino. 
Their effect on clustering is thus partially degenerate with the effect of dark forces \cite{Archidiacono:2022iuu,Bottaro:2023wkd,2407.18252,Green:2024xbb,Graham:2025dqn,Costa:2025kwt}. 
Posterior distributions on $\sum m_\nu$ and $\beta$ obtained from the baseline dataset \textit{Planck} + ACT\textit{lens} + DESI DR1 BAO+FS are shown in Fig.~\ref{fig:nus}. 
Upon freeing the neutrino masses, the strength of the dark fifth force remains approximately the same as before, with $\beta=0.0046 \pm 0.0022$ albeit with a roughly 30\% larger errorbar (see also \cite{Costa:2025kwt}). 
Accordingly, the bound on the total neutrino mass is relaxed to $\sum m_\nu < 0.19 \, e\textrm{V}$ at $95\%$CL, compared to $\sum m_\nu \lesssim 0.07 \, e\textrm{V}$ in a $\Lambda$CDM Universe~\cite{2503.14744,2503.14738,2411.12022}. 
Our result is very similar to the one presented in Ref.~\cite{Costa:2025kwt}, which uses BAO DR2 data but does not include any full-shape information, therefore indicating that the relaxation of the bound is mostly of geometric nature. It will be interesting to reassess these findings once the DESI DR2 catalogs become available and a full-shape analysis will be possible.

\section{Conclusions}
\label{sec:conclusions}
%==============================================================

Cosmological data provide a direct test of long-range dynamics in the Dark Sector.
In this work we have used recent DESI measurements, in combination with CMB data, to test whether the mild tension between low-redshift LSS observations and the CMB-inferred $\Lambda$CDM cosmology can be interpreted as evidence for a violation of the Equivalence Principle by dark matter.
The physical effect is twofold: the dark matter density redshifts faster than in $\Lambda$CDM, modifying the late-time expansion history, while the attractive fifth force enhances the growth of matter perturbations.

We find that \textit{Planck} primaries + ACT lensing data, combined either with DESI DR2 BAO or with DESI DR1 FS and BAO measurements, favor the presence of a fifth force with strength of $\beta \sim 5\text{‰}$ relative to gravity at the $\sim 2.5\sigma$ level.
This preference is strengthened to $\sim 3.1\sigma$ in a preliminary combination of FS DR1 with BAO DR2 data, although the result should be interpreted with care since the cross-correlation between the two datasets is ignored.
Allowing neutrino masses to vary  while weakening the preference for a non-zero value of $\beta$ yields a $2\sigma$-bound on $\sum m_\nu < 0.19 \, e \textrm{V}$ consistent with minimal neutrino mass determined by oscillation experiments.

These results suggest that current LSS data are already sensitive to dark forces at the permille level relative to gravity, and that the DESI--CMB tension may be pointing toward new dynamics in the Dark Sector. We have  not yet included several additional probes, like cosmic shear, galaxy-galaxy lensing, and cross-correlations of the CMB lensing maps with galaxy catalogs, which are expected to bring in additional constraining power.
Finally, upcoming DESI DR2 full-shape data and complementary measurements from ACT and SPT primary CMB data, and \textit{Euclid}~\cite{Euclid:2024yrr} in the near future will further tighten the constraints and clarify this interpretation.

\acknowledgments
We thank Ennio Salvioni, Diego Redigolo and Francesco Verdiani for helpful discussions. 
The numerical results presented in this work made extensive use of public softwares including \texttt{CLASS}~\cite{Blas:2011rf,Lesgourgues:2011re}, \texttt{PyBird}~\cite{DAmico:2020kxu}, \texttt{Cobaya}~\cite{Torrado:2020dgo}, \texttt{GetDist}~\cite{Lewis:2019xzd} and \texttt{Prospect}~\cite{Holm:2023uwa}.
Some computations were performed on the  INDACO high-performance computing platform at the University of Milan. 
PZ acknowledges support from Fondazione Cariplo under the grant No 2023-1205.

This research used data obtained with the Dark Energy Spectroscopic Instrument (DESI). DESI construction and operations is managed by the Lawrence Berkeley National Laboratory. This material is based upon work supported by the U.S. Department of Energy, Office of Science, Office of High-Energy Physics, under Contract No. DE–AC02–05CH11231, and by the National Energy Research Scientific Computing Center, a DOE Office of Science User Facility under the same contract. Additional support for DESI was provided by the U.S. National Science Foundation (NSF), Division of Astronomical Sciences under Contract No. AST-0950945 to the NSF’s National Optical-Infrared Astronomy Research Laboratory; the Science and Technology Facilities Council of the United Kingdom; the Gordon and Betty Moore Foundation; the Heising-Simons Foundation; the French Alternative Energies and Atomic Energy Commission (CEA); the National Council of Humanities, Science and Technology of Mexico (CONAHCYT); the Ministry of Science and Innovation of Spain (MICINN), and by the DESI Member Institutions: www.desi.lbl.gov/collaborating-institutions. The DESI collaboration is honored to be permitted to conduct scientific research on I’oligam Du’ag (Kitt Peak), a mountain with particular significance to the Tohono O’odham Nation. Any opinions, findings, and conclusions or recommendations expressed in this material are those of the author(s) and do not necessarily reflect the views of the U.S. National Science Foundation, the U.S. Department of Energy, or any of the listed funding agencies.

\IfFileExists{JHEP.bst}{\bibliographystyle{JHEP}}{\bibliographystyle{../JHEP}}
%\IfFileExists{body/5th_desi.bib}{\bibliography{body/5th_desi}}{\bibliography{5th_desi}}

\bibliography{5th_desi_references}

\clearpage
\newpage
\appendix
\onecolumngrid
\clearpage
\newpage

%======================================================
\section{Details on the cosmological inference setup}
\label{sec:desi_dr1_full_shape}
%=====================================================

\paragraph{Inference setup}
To infer the posterior distribution of physical parameters of interest from cosmological data described in Sec.~\ref{sec:results}, we perform Markov-Chain Monte-Carlo (MCMC) sampling using the Metropolis-Hastings algorithm implemented in the \texttt{Cobaya} cosmological inference framework \cite{Torrado:2020dgo}.\footnote{\url{https://github.com/CobayaSampler/cobaya}} 
To analyze the combination of various datasets we  sample the product of their likelihoods together with the prior specified in Table~\ref{tab:desi_pybird_priors}. 
In general, we vary the following set of cosmological parameters: $\lbrace \beta, \Omega_{\chi}, \omega_{\rm b}, \log(10^{10}A_s), n_s, \theta_s, \tau_{\rm reio} \rbrace$, corresponding respectively to the dark fifth force relative strength to gravity, the today's interacting dark matter abundance, the today's baryons abundance, the log-amplitude of the primordial fluctuations, the spectral tilt, the angular size of the sound horizon at recombination, and the reionization optical depth. 
Our results are instead mainly presented in terms of (derived) parameters $\lbrace \beta, \Omega_{\rm m}, H_0\rbrace$, where the today's total matter density is $\Omega_{\rm m} = \Omega_\chi + \Omega_{\rm b}$, since in this work all DM is considered to be interacting. 
We consider three degenerate massive neutrinos with their total mass fixed to minimal in normal hierarchy, $\sum m_\nu \equiv 0.06 \, e \textrm{V}$, except if specified otherwise. 
The background and linear cosmological quantities are computed with our modified version of the Boltzmann solver~\texttt{CLASS}~\cite{Blas:2011rf,Lesgourgues:2011re,Archidiacono:2022iuu},\footnote{\url{http://class-code.net/}} while the nonlinear galaxy power spectrum model to DESI FS data is computed with~\texttt{PyBird}~\cite{DAmico:2020kxu}.\footnote{\url{https://github.com/pierrexyz/pybird}}
Sampling convergence is monitored via the Gelman-Rubin criterion. 
All sampled posterior distributions are inferred from 10 MCMC chains run in parallel, satisfying $R-1 \lesssim 0.5\, \%$ convergence on all sampled parameters. 
Global best fits (\textit{i.e.},~maximum a posteriori estimates) are found via simulated annealing implemented in the profile-likelihood code \texttt{Prospect} \cite{Holm:2023uwa}. 
In the following, we provide further details on our analysis of DESI DR1 data. 

\begin{table*}[ht!]
\centering

    %\renewcommand{\arraystretch}{1.08}

    % Left panel
    \begin{minipage}[t]{0.48\textwidth}
        \centering

        \setlength{\tabcolsep}{5pt}
        \begin{tabular*}{\linewidth}{
            @{\extracolsep{\fill}} llc @{}
        }
            \toprule
             & Parameter & Prior \\
            \midrule

            \multirow{7}{*}{Cosmological parameter}
            & $\beta$
            & $\mathcal{U}(0,0.045)$ \\

            & $\Omega_{\chi}$
            & $\mathcal{U}(0,1)$ \\

            & $n_s$
            & $\mathcal{U}(0.8,1.2)$ \\

            & $\log(10^{10}A_s)$
            & $\mathcal{U}(1.61,3.91)$ \\

            & $100\,\theta_s$
            & $\mathcal{U}(0.5,10)$ \\

            & $\omega_{\rm b}$
            & $\mathcal{U}(0.005,0.1)$ \\

            & $\tau_{\rm reio}$
            & $\mathcal{U}(0.01,0.8)$ \\

            \bottomrule
        \end{tabular*}
    \end{minipage}
    \hfill
    % Right panel
    \begin{minipage}[t]{0.49\textwidth}
        \centering

        \setlength{\tabcolsep}{5pt}
        \begin{tabular*}{\linewidth}{
            @{\extracolsep{\fill}} llc @{}
        }
            \toprule
             & Parameter ($i = 1, \dots, 6$) & Prior \\
            \midrule

            \multirow{3}{*}{Galaxy bias}
            & $b_{1,i}$
            & $\mathcal{U}(0.5,3)$ \\

            & $c_{2,i}$
            & $\mathcal{N}(0,5^2)$ \\

            & *$b_{3,i}$
            & $\mathcal{N}(0,5^2)$ \\

            \midrule

            \multirow{3}{*}{Counterterm}
            & *$c_{{\rm ct},i}$
            & $\mathcal{N}(0,5^2)$ \\

            & *$c_{r1,i}$
            & $\mathcal{N}(0,5^2)$ \\

            & *$c_{r2,i}$
            & $\mathcal{N}(0,5^2)$ \\

            \midrule

            \multirow{3}{*}{Stochastic term}
            & *$c_{\epsilon0,i}$
            & $\mathcal{N}(0,2^2)$ \\

            & *$c_{\epsilon1,i}$
            & $\mathcal{N}(0,5^2)$ \\

            & *$c_{\epsilon2,i}$
            & $\mathcal{N}(0,5^2)$ \\

            \bottomrule
        \end{tabular*}
    \end{minipage}

\caption{
Priors on cosmological and nuisance parameters used for all combined CMB + LSS data analyses presented in this work (except if specified otherwise). 
Parameters are either varied within a flat prior {\small$\mathcal{U}(a,b)$} or a Gaussian prior {\small$\mathcal{N}(\mu,\sigma^2)$}.  
Cosmological parameters (\textit{left panel}) are scanned along nuisance parameters entering the CMB likelihoods (\textit{not shown}) or our DESI DR1 FS likelihood (\textit{right panel}). 
All EFT parameters are unitless, as the counterterms are normalized by $k_{\rm M}^2$ and $k_{\rm R}^2$ for $c_{\textrm{ct},i}$ and $c_{r1,i}$, $c_{r2,i}$, respectively, with $k_{\rm M} / k_{\rm R} = 0.7 / 0.25 \, h\, \textrm{Mpc}^{-1}$, and the stochastic terms by the relevant mean number density $\bar n$
for each tracer $i$, wherein the scale dependent ones $c_{\epsilon1,i}$ and $c_{\epsilon2,i}$ are further normalized by $k_{\rm M}^{2}$. 
Parameters marked with a * enter at most quadratically in the likelihood and are thus analytically marginalized, while all other parameters are scanned in the numerical sampling. }
\label{tab:desi_pybird_priors}
\end{table*}

\begin{figure*}[ht!]
\centering
\begin{minipage}[c]{0.50\textwidth}{
\begin{tabular}{lccc}
\toprule
Parameter & DESI 2024~\cite{2411.12021} & This work & $\,\,\Delta \bar p/\sigma$ \\
\midrule
$+0.14$ \\
$\Omega_m$ & $0.296\pm0.010$ & $0.296\pm0.011$ & $\,\,-0.03$ \\
$H_0\,[{\rm km}\,{\rm s}^{-1}{\rm Mpc}^{-1}]$ & $68.63\pm0.78$ & $68.67\pm0.79$ & $\,\,+0.05$ \\
$\sigma_8$ & $0.842\pm0.034$ & $0.809\pm0.034$ & $\,\,-0.96$ \\
$n_s$ & $0.995\pm0.028$ & $0.983\pm0.030$ & $\,\,-0.42$  \\
\bottomrule
\end{tabular}
}
\end{minipage}
\hspace{0.005\textwidth}
\begin{minipage}[c]{0.48\textwidth}
\vspace{-23pt}
\centering
\includegraphics[width=\linewidth]{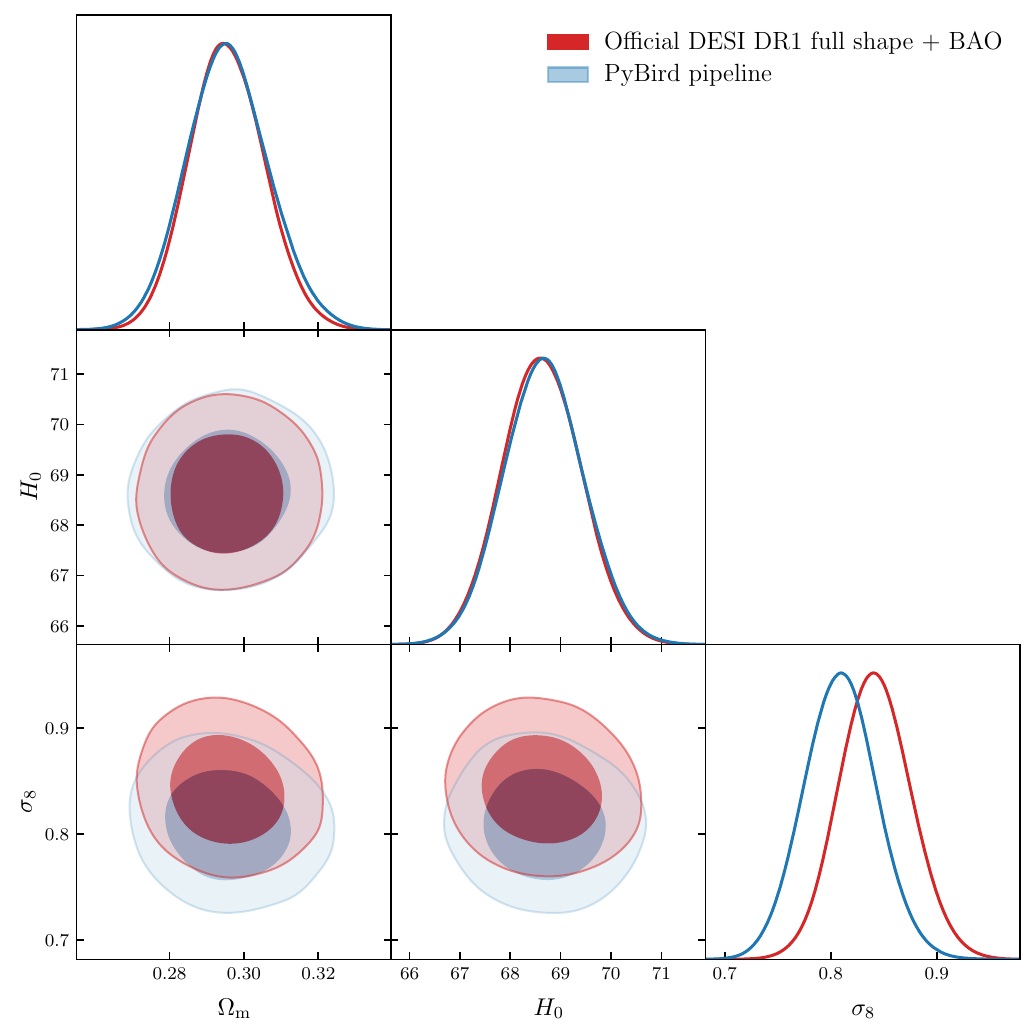}
\vspace{-14pt}
\end{minipage}
\caption{Comparison on $\Lambda$CDM parameters inferred from DESI DR1 FS+BAO between DESI official pipeline~\cite{2411.12021} and our \textit{PyBird}--\textit{Cobaya} likelihood, obtained with a BBN prior on $\omega_{\rm b} \sim \mathcal{N}(0.02218, 0.00055^2)$ and $\sim \mathcal{N}(0.02198, 0.00053^2)$, respectively, and a $10\, \times$\textit{Planck} prior on $n_s \sim \mathcal{N}(0.965, 0.042^2)$. 
\textit{Left panel}: $68\%$-credible intervals (mean and standard deviation) and relative mean differences in units of the average standard deviation, $\sigma \equiv (\sigma_{\rm DESI}+\sigma_{\rm ours})/2$.
\textit{Right panel}: 1D and 2D posterior distributions. }
\label{fig:validate_desi_pipeline_lcdm}
\end{figure*}

\paragraph{DESI DR1 FS+BAO likelihood analysis} We analyze the public DESI DR1 galaxy and quasar power-spectrum measurements \cite{2411.12020,2411.12021} in the six following tracer/redshift bins: BGS, LRG1, LRG2, LRG3, ELG2, and QSO.
Following the DESI DR1 FS baseline analysis \cite{2411.12021}, for each bin $i=1,\dots, 6$, we use the redshift-space monopole and quadrupole, $\ell=0,2$, in the range $0.02 \leq k/[h\,{\rm Mpc}^{-1}] \leq 0.20$. 
The theoretical modelling is based on the EFTofLSS \cite{Baumann:2010tm,Carrasco:2012cv}, and its formulation for biased tracers in redshift space \cite{Perko:2016puo}. 
The galaxy-clustering preditions and likelihood are evaluated with the \texttt{PyBird} code \cite{DAmico:2020kxu,Reeves:2025bxn,Lai:2024bpl}, embedded in \texttt{Cobaya}. 

\texttt{PyBird} computes the one-loop EFTofLSS prediction for biased tracers in redshift space, including the Alcock-Paczynski (AP) mapping, IR resummation, and the convolution with the survey window function.
Schematically, for each redshift bin, \texttt{PyBird} computes
\begin{equation} \label{eq:pybird_decomposition}
    P^{\rm th}_{g,\ell}(k)
    =
    P^{\rm lin}_{g,\ell}(k)
    +
    P^{\rm 1loop}_{g,\ell}(k)
    +
    P^{\rm ct}_{g,\ell}(k)
    +
    P^{\rm stoch}_{g,\ell}(k)\,,
\end{equation}
where the first two terms are the tree-level and one-loop biased-tracer contributions, $P^{\rm ct}_{g,\ell}$ denotes the EFT counterterms, and $P^{\rm stoch}_{g,\ell}$ denotes the stochastic contributions.
We use the \texttt{PyBird} `westcoast' bias basis \cite{DAmico:2020kxu,Simon:2022lde}, in which the bias sector is parametrized by $\{b_1,c_2,c_4,b_3\}$, while the counterterms and stochastic parameters are parametrized by $\{c_{\rm ct},c_{r1},c_{r2}\}$ and $\{c_{\epsilon 0},c_{\epsilon 1},c_{\epsilon 2}\}$.
The quadratic bias parameter $c_4$ is fixed to zero in our analyses as found to be unconstrained by the one-loop power spectrum~\cite{DAmico:2019fhj,Nishimichi:2020tvu}, while the remaining nuisance parameters are either directly sampled or analytically marginalized, as reported in Table~\ref{tab:desi_pybird_priors}.
The counterterms and stochastic terms are normalized with the EFT scales $k_{\rm M} = 0.7\,h\,{\rm Mpc}^{-1}$, $k_{\rm R} = 0.25\,h\,{\rm Mpc}^{-1}$~\cite{DAmico:2021ymi}. 
The stochastic terms are further normalized by the mean number density $\bar n_i$ of each DESI redshift bin~\cite{2411.12020}. 
Within the adopted scale cut, no sizable theoretical error is introduced beyond the one-loop corrections, EFT counterterms, and stochastic parameters already marginalized in the likelihood, as validated on simulations~\cite{Maus:2024sbb,Lai:2024bpl,Nishimichi:2020tvu}. 

The fifth-force cosmology enters the DESI likelihood through the linear total matter power spectrum and the linear growth rate returned by the modified version of the Boltzmann solver \texttt{CLASS}~\cite{Blas:2011rf,Lesgourgues:2011re,Archidiacono:2022iuu}. 
More explicitly, at each effective redshift $z_i$, \texttt{PyBird} receives the background and linear perturbation quantities $\{P_{\rm lin}(k,z_i),\,f(z_i),\,H(z_i),\,D_A(z_i)\}$, and returns the nonlinear -- window-convolved, AP-corrected -- galaxy power spectrum multipoles of~\eqref{eq:pybird_decomposition}. 
This implementation captures the leading logarithmically enhanced effects of the long-range dark force on the total matter perturbations.
Following the EFTofLSS treatment of dark fifth forces \cite{Bottaro:2023wkd}, we do not introduce additional independent bias operators in the baseline power-spectrum likelihood, which are subleading at the level of the two-point function for the small values of $\beta$ considered in this work (see Sec.~\ref{sec:theory} for more details).

We also consider combination of DESI DR1 FS with BAO data from DR1 or DR2~\cite{2404.03000,2503.14738}. 
At each effective redshift $z_i$, the theory prediction for the BAO sector is written in terms of the standard transverse and radial distance ratios, computed from $D_M(z_i)$, $D_H(z_i)$, and $r_{\rm d}$. 
The BAO measurements are combined with the corresponding FS data vector using the covariance distributed with the DESI DR1 likelihood, estimated from 1000 EZmock realizations per tracer -- including all systematic and Hartlap factor corrections as described in ref.~\cite{2411.12021}. 
To combine DR1 FS with DR2 BAO, we instead simply do  the product of their individual likelihoods, neglecting their cross-correlation. 
The final DESI likelihood is Gaussian,
\begin{equation}\label{eq:desi_gaussian_likelihood}
    -2\ln \mathcal{L}_{\rm DESI}
    =
    \big(\mathbf{D}-\mathbf{T}\big)^{\!T}
    \mathbf{C}^{-1}
    \big(\mathbf{D}-\mathbf{T}\big)
    -2\ln \Pi_{\rm EFT}\,,
\end{equation}
where $\mathbf{D}$ is the DESI data vector consisting of the concatenated power spectrum multipoles with eventually the BAO distance measurements, $\mathbf{T}$ is the corresponding theory prediction, $\mathbf{C}$ is the DESI public covariance matrix estimated with mock catalogs \cite{Forero-Sanchez:2024bjh}, and $\Pi_{\rm EFT}$ denotes the product of the Gaussian EFT priors specified in Tab.~\ref{tab:desi_pybird_priors}.
%The EFT parameters that enter linearly in the model are analytically marginalized inside the PyBird likelihood.
In DESI DR1 all tracers are assumed to be independent, such that each tracer data can be described by an independent Gaussian likelihood of the form of~\eqref{eq:desi_gaussian_likelihood}, from which the full DESI DR1 likelihood is simply obtained by their product. 

We validate in Fig.~\ref{fig:validate_desi_pipeline_lcdm} our implementation by comparing $\Lambda$CDM limits obtained by fitting DESI DR1 FS+BAO (no Lyman-$\alpha$) from our \texttt{PyBird}--\texttt{Cobaya} pipeline with the public posterior chain from DESI~\cite{2411.12021}.\footnote{Available from the \href{https://data.desi.lbl.gov/doc/releases/dr1/vac/full-shape-cosmo-params/}{DESI DR1 data release}.}
We find an average absolute shift on the cosmological parameters of $0.36 \,\sigma$, with the main shift in the direction of the clustering amplitude $\sigma_8$, of about $0.96\,\sigma$.
Since the amplitude of the primordial fluctuations as measured by LSS plays a minor role once combined with CMB data and is little correlated with $\beta$, we conclude that residual differences between our implementation and the official DESI analysis are subdominant for the fifth-force constraints reported in this work.

%===========================================
\section{Full cosmological triangle plots}
\label{app:constraints}
%=============================================

\begin{figure}[h!]
    \centering
    \includegraphics[width=.9\linewidth]{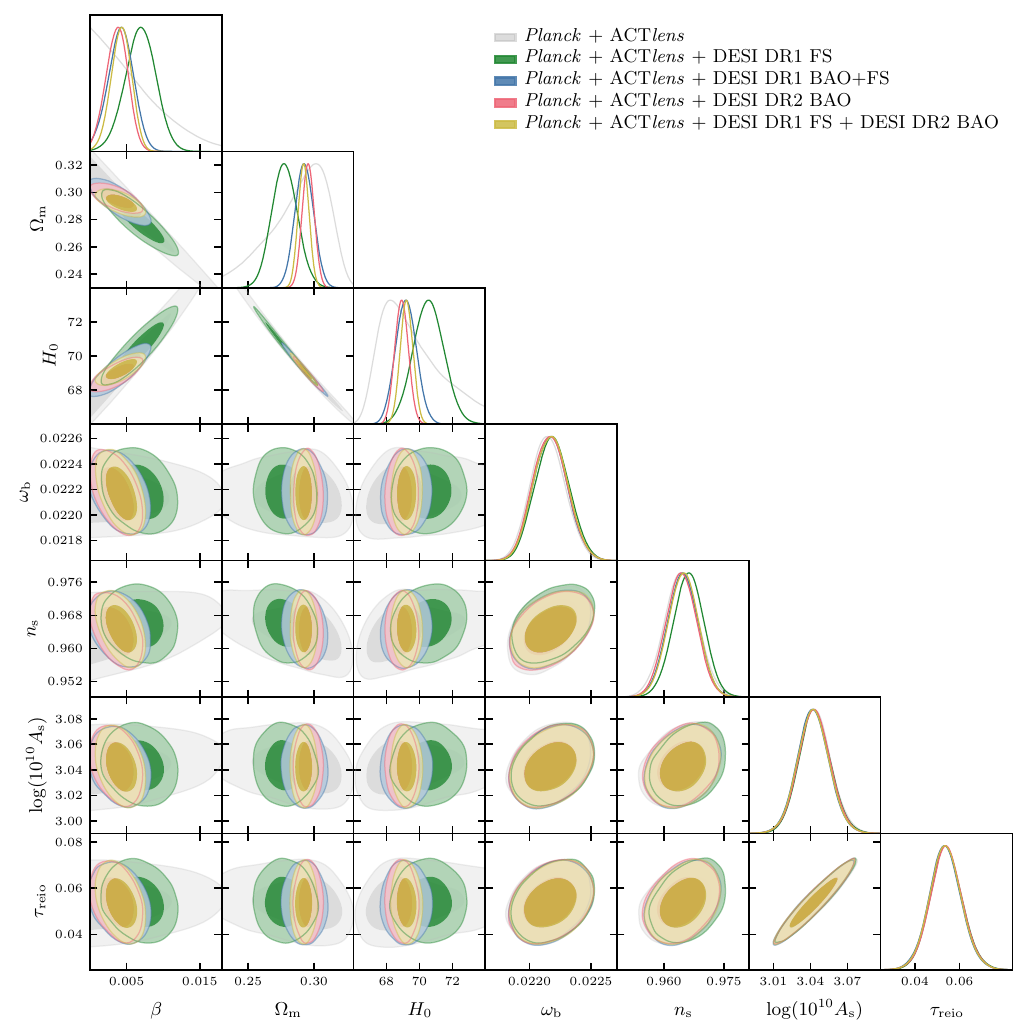}

    \vspace{-10pt}
    \caption{1D and 2D posterior distributions of the full set of cosmological parameters, for the various dataset combinations between CMB and LSS probes considered in this work. }
    \label{fig:posterior_all}
\end{figure}

In Fig.~\ref{fig:posterior_all}, we show the 1D and 2D posterior distributions of the full set of cosmological parameters for the various dataset combinations considered in this work.

\onecolumngrid

\end{document}